\documentclass[12pt]{article}
\usepackage[left=2.5cm,top=2.50cm,right=2.5cm,bottom=2.50cm]{geometry}
\usepackage{mathrsfs}
\usepackage{amsmath,amssymb,latexsym,color,cancel,graphicx,bbm,colortbl}
\usepackage[english]{babel}
\usepackage[latin1]{inputenc}
\usepackage{ragged2e}
\begin{document}
\date{}

\title{ $SU(1,1)$ Approach to Stokes Parameters and the Theory of  Light  Polarization }

\author{R. D. Mota$^{a}$, D. Ojeda-Guill\'en$^{b}$, M. Salazar-Ram{\'i}rez$^{b}$ \\ and V. D. Granados$^{c}$} \maketitle

\begin{minipage}{0.9\textwidth}

\small $^{a}$ Escuela Superior de Ingenier\'ia Mec\'anica y
El\'ectrica, Unidad Culhuacan, IPN.  Av. Santa Ana No. 1000, Col. San
Francisco Culhuacan, Delegaci\'on Coyoacan C.P.04430,  Ciudadad de M\'exico , Mexico.\\

\small $^{b}$Escuela Superior de C\'omputo, Instituto Polit\'ecnico Nacional, Av. Juan de Dios Batiz, Esq. Av. 
Miguel Othon de Mendiz\'abal, Col. Lindavista, Del. Gustavo A. Madero, C.P. 07738,  Ciudad de M\'exico, Mexico.\\

\small $^{c}$ Escuela Superior de F\'{\i}sica y Matem\'aticas,
Instituto Polit\'ecnico Nacional,
Ed. 9, Unidad Profesional Adolfo L\'opez Mateos, 07738, Ciudad de M\'exico, Mexico.\\
\end{minipage}

\abstract{We introduce an alternative approach to the polarization theory of light.
This is based on a set of quantum operators, constructed from two independent 
bosons, being three of them the $su(1,1)$ Lie algebra generators, and the other one, the Casimir
operator of this algebra. By taking the expectation value of these generators in a two-mode coherent
state, their classical limit is obtained. We use these classical quantities to define the new Stokes-like 
parameters.  We show that the light polarization ellipse can be written 
in terms of the Stokes-like parameters. Also, we write these parameters in terms of other two 
quantities, and show that they define a one-sheet  (Poincar\'e hyperboloid) of a two-sheet hyperboloid. 
Our study is restricted to the case of a monochromatic plane electromagnetic wave which propagates
along the $z$ axis.}\\
\linebreak

\maketitle

\section{Introduction}
G. Stokes studied the polarization properties of a
quasi-monochromatic plane wave of light in an arbitrary
polarization state by introducing four quantities, known since then as
the Stokes parameters \cite{stokes}.

The Stokes parameters are a set of four quantities which provide 
intuitive and practical tools to describe the polarization of light \cite{sur}.
The stokes parameters  give a direct relation between the light (photons) polarization 
and the polarization of elementary particles \cite{mc1}. This fact was widely exploited
to study many features of radiation of particles and to 
the scattering problems \cite{mc2,colet}. 
   
The Stokes parameters are formulated in terms of the observables of 
the electromagnetic field, namely, the amplitudes and the relative phase 
difference between the orthogonal components of the field \cite{wolf}. In fact, the 
density matrix \cite{mc1} and the coherence matrix \cite{wolf59} for the case
of electromagnetic radiation result to be the same \cite{blaskal-kim}, and are written 
in terms of these observables.

The standard procedure to describe the polarization of an electromagnetic wave is to set the 
propagation direction along the $z$-axis, and the two components of the
polarization field on the $x-$ and $y-$directions. However,            
when the direction of arrival from the source is unknown a
priori, the three-dimensional coherence matrix must be used to
obtain a complete polarization characterization \cite{roman,rama,carozzi,mota}.  
In Ref. \cite{jauch}, Jauch and Rohrlich 
introduced the stokes parameters in the quantum regime, which
are called Stokes operators. It is at the quantum domain where we can see that a 
symmetry group structure is related to the Stokes operators. When the direction of 
propagation of light is known, the symmetry is the $SU(2)$ group \cite{jauch,canadian}. However,
when the direction of propagation is unknown, the symmetry group is $SU(3)$ 
\cite{roman,rama,carozzi,mota}. Also, other generalizations of Stokes operators  have been reported  
\cite{sergienko,jaeger}.

In this work we give a new approach to the theory of 
light polarization. For simplicity, we study the case of a monochromatic plane electromagnetic
wave which propagates along the $z$-axis.
Our study is based on a set of quantum operators, constructed from two independent 
bosons, being three of them the $su(1,1)$ Lie algebra generators, and the other one, the Casimir
operator of this algebra.
This work is organized as follows. In section 2, we deduce the $su(1,1)$ Lie algebra generators  by 
the Jordan-Schwinger map. In section 3,  by taking the
expectation value of the algebra generators in a two-mode coherent
state, we obtain their classical limit.  In section 4, we define our Stokes parameters (we refer to them 
Stokes-like parameters)  and show that the light polarization ellipse can be 
written in terms of them. In section 5, the Stokes-like parameters are written in terms of two parameters  
and it is shown that they define  a one-sheet (Poincar\'e hyperboloid) of a two-sheet hyperboloid. 

\section{ The Jordan-Schwinger map and the $su(1,1)$ Lie algebra generators}

In what follows we will use $\hbar=\mu= 1$, where $\mu$ is
the mass of each one-dimensional harmonic oscillator and $\omega$
is the frequency of either the electromagnetic wave or the
harmonic oscillators.

We define the operators 
\begin{equation}
\mathcal{K}_i=B^{\dag}\frac{\Sigma_i}{2}B \hspace{5ex} i=0,1, 2, 3\label{JSM}
\end{equation}
with
\begin{equation}
 B=\begin{pmatrix}a_1\\
    a_2^\dag\end{pmatrix},   \hspace{5ex}      B^\dag=(a_1^\dag, a_2).
\end{equation}
The operators $a_1$ and $a_2$ the left and right annihilation operators of the two-dimensional harmonic
oscillator, with the non vanishing  commutators $[a_1,a_1^{\dag}]=[a_2,a_2^{\dag}]=1$.  
The matrices $\Sigma_i$ are defined as follows: $\Sigma_0=\sigma_3$, $\Sigma_1=\sigma_1$, $\Sigma_2=\sigma_2$ and $\Sigma_3=1_{2\times 2}$, where $\sigma_i$ are the usual Pauli matrices \cite{cohen}.  \\
Explicitly, the operators $\mathcal{K}_0$, $\mathcal{K}_1$, $\mathcal{K}_2$ and $\mathcal{K}_3$ are
given by
\begin{eqnarray}
&&\mathcal{K}_0=\frac{1}{2}(a_1^\dag a_1-a_2 a_2^\dag ),\hspace{8ex}\mathcal{K}_1=\frac{1}{2}(a_1^\dag a_2^\dag+a_2 a_1)\label{ST1}\\
&&\mathcal{K}_2=\frac{i}{2}(-a_1^\dag a_2^\dag+a_2 a_1),\hspace{6ex}\mathcal{K}_3=\frac{1}{2}(a_1^\dag a_1+a_2 a_2^\dag)\label{ST2}.
\end{eqnarray}
The operator $\mathcal{K}_0$ can be rewritten as $\mathcal{K}_0=\frac{1}{2}(a_1^\dag a_1-a_2^\dag a_2-1)\equiv \frac{1}{2}(\mathcal{L}_z-1)$, being $\mathcal{L}_z$ the $z-$component of the angular momentum of the two-dimensional harmonic oscillator, whose Hamiltonian is given by $H=\omega(a_1^\dag a_1+a_2^\dag a_2 +1)$. Therefore, the operator $\mathcal{K}_3$ is essentially the  energy of the two-dimensional harmonic oscillator. It can be shown that 
\begin{equation}
[\mathcal{K}_0,H]=0, \hspace{2ex} 
\end{equation}
Also, by a straightforward calculation, we show that the commutation relations of the operators $\mathcal{K}_i$ are
\begin{equation}
\left[\mathcal{K}_1,\mathcal{K}_2\right]=-i \mathcal{K}_3,\hspace{3ex} \left[\mathcal{K}_2,\mathcal{K}_3\right]=i \mathcal{K}_1 ,\hspace{3ex} \left[\mathcal{K}_3,\mathcal{K}_1\right]=i \mathcal{K}_3.
\label{comu}
\end{equation}
Therefore, these operators close the $su(1,1)$ Lie algebra. The Casimir operator $C\equiv \mathcal{K}_3^2-\mathcal{K}_2^2-\mathcal{K}_1^2$ for 
this algebra results to be  $C=\mathcal{K}_0(\mathcal{K}_0+1)$. Hence, we have obtained the generators of the $su(1,1)$ Lie algebra,  equation  (\ref{JSM}), by the so called Jordan-Schwinger map \cite{bieden}. 

\section{Classical limit of the $SU(1,1)$ generators} 

 In Ref. \cite{cohen} the classical states of the two-dimensional harmonic oscillator were calculated. On the other 
hand, the stokes parameters were obtained by evaluating the $su(2)$ Lie algebra generators in a two-mode coherent state \cite{canadian}. The same idea was used to derive one of the Stokes parameters generalizations by evaluating the $su(3)$ Lie algebra generators in a three-mode coherent states \cite{mota}. In this paper we take advantage of these facts. Thus, we 
evaluate the $su(1,1)$ Lie algebra generators ${\mathcal{K}_i}$ in a two-mode coherent state to obtain their classical limit.
The two-mode coherent states are well known to be  given by      
\begin{equation}
|\alpha_1,\alpha_2\rangle=\exp{\left[-\frac{1}{2}\left(|\alpha_1|^2+|\alpha_2|^2\right)\right]}\sum_{n_1,n_2=0}^\infty \frac{\alpha_1^{n_1}\alpha_2^{n_2}}
{\sqrt{n_1!n_2!}}|n_1,n_2\rangle,\label{TM}
\end{equation}
which are eigenstates of the annihilation operators $a_1$ and $a_2$  
\begin{equation}
a_i |\alpha_1,\alpha_2\rangle=\alpha_i|\alpha_1,\alpha_2\rangle.
\end{equation}
With this procedure, we obtain 
\begin{align}
\langle \mathcal{K}_0\rangle_\alpha &= \frac{1}{2}\left(|\alpha_{1}|^2-|\alpha_{2}|^2\right),&
\langle \mathcal{K}_1\rangle_\alpha &= \frac{1}{2}\left(\alpha_{1}^*\alpha_{2}^*+ \alpha_{1}\alpha_{2}\right),\nonumber\\
\langle \mathcal{K}_2\rangle_\alpha &= \frac{i}{2}\left(\alpha_{1}\alpha_{2}-\alpha_{1}^*\alpha_{2}^*\right), &
\langle \mathcal{K}_3\rangle_\alpha &= \frac{1}{2}\left(|\alpha_{1}|^2+|\alpha_2|^2\right),
\end{align}
where $\alpha_i$ are the classical oscillations $\alpha_i=|\alpha_{0i}|\exp[-i(\omega t-\phi_i)]$ with amplitudes 
$|\alpha_{0i}|$ and phases $\phi_i$. This leads to
\begin{align}
\langle \mathcal{K}_0\rangle_\alpha &= \frac{1}{2}\left(|\alpha_{01}|^2-|\alpha_{02}|^2\right),&
\langle \mathcal{K}_1\rangle_\alpha &= |\alpha_{01}||\alpha_{02}|\cos(2\omega t -\sigma_{21}),\label{funda1}\\
\langle \mathcal{K}_2\rangle_\alpha &= |\alpha_{01}||\alpha_{02}|\sin(2\omega t -\sigma_{21}) , &
\langle \mathcal{K}_3\rangle_\alpha &= \frac{1}{2}\left(|\alpha_{01}|^2+|\alpha_{02}|^2\right)\label{funda2}.
\end{align}
where $\sigma_{21}\equiv \phi_2+\phi_1$.

Thus, the classical limit of the Lie algebra generators in a time-dependent two-mode coherent state
is time dependent. This is because the $su(1,1)$ Lie algebra generators do not commute with the 
Hamiltonian of the two-dimensional harmonic oscillator.  It is well known that the standard Stokes
parameters obtained as the classical limit of the $su(2)$ generators are time-independent \cite{canadian}.
This is due to the $su(2)$ Lie algebra generators commute with the two-dimensional harmonic oscillator Hamiltonian.  

\section{Polarization ellipse for an electromagnetic wave and the Stokes-like parameters}

Our procedure is  based on the experience gained in deducing the 
time-independence of the polarization ellipse for the superposition 
of two monochromatic electromagnetic waves, as we can be see, 
for example,  in Ch. 2 of Ref. \cite{guenther} or in Ch. 8 of Ref. \cite{hecht}.  
We notice that in these deductions the reason for this
time-independence rests on the mathematical properties of the trigonometric 
functions and that we are treating with monochromatic waves. Also, in the 
case when the amplitudes and phases fluctuate slowly (the so-called 
quasi-monochromatic or nearly  monochromatic case) compared 
to the rapid vibrations of the sinusoid and the cosinusoid functions, 
it  can be derived the same polarization ellipse \cite{colet}. 
In this case, the amplitudes and phases change slowly with time.
However, in the present work we are considering monochromatic 
waves only.

We set an electromagnetic wave with arbitrary polarization, which propagates along the $z$-axis, given by
\begin{equation}
\vec{E}(z,t)=(\mathbf{ i}\alpha_1+\mathbf{j}\alpha_2)e^{i k z}+ C. C.
\end{equation}
where the complex amplitudes are defined by  $\alpha_i=|\alpha_{0i}|\exp{[-i(\omega t -\phi_i)]}$. It can be written in the form
\begin{equation}
\vec{E}(z,t)=E_1\mathbf{i}+E_2\mathbf{j}\label{campo},
\end{equation}
where   
\begin{align}
E_1&= A_1\cos(\omega t-kz)+B_1\sin(\omega t-kz)\label{campo1},\\
E_2&=  A_2\cos(\omega t-kz)+B_2\sin(\omega t-kz)\label{campo2},
\end{align}
and 
\begin{equation}
A_i=2|\alpha_{0i}|\cos{\phi_i} \hspace{3ex}B_i=2|\alpha_{0i}|\sin{\phi_i}\hspace{3ex} i=1,2.
\end{equation}
From these equations it  is easy to obtain the nonparametric quadratic equation
\begin{eqnarray}
E_1^2(A_2^2+B_2^2)+E_2^2(A_1^2+B_1^2)
-2E_1E_2(A_1A_2+B_1B_2)\nonumber\\
=(A_1B_2-A_2B_1)^2.
\label{trayec}
\end{eqnarray}
By setting  the definitions $\mathcal{A}=A_2^2+B_2^2$, $\mathcal{B}=-2(A_1A_2+B_1B_2)$ and
$\mathcal{C}=A_1^2+B_1^2$, it is shown that
$\mathcal{B}^2-4\mathcal{AC}=-4(A_1B_2-A_2B_1)^2$. Thus, when $A_1B_2\neq A_2B_1$, equation (\ref{trayec}) 
represents an ellipse centered
at the origin.
Equations (\ref{campo1}) and (\ref{campo2}) allow to write equation (\ref{trayec}) as
\begin{eqnarray}
E_1^2|\alpha_{02}|^2+E_2^2|\alpha_{01}|^2-E_1 E_2|\alpha_{01}||\alpha_{02}|\cos(\phi_2-\phi_1)\nonumber\\
=2|\alpha_{01}|^2|\alpha_{02}|^2\sin^2(\phi_2-\phi_1).
\end{eqnarray}
This is the polarization ellipse described by the plane wave electromagnetic field.   
Using equations (\ref{funda1}) and (\ref{funda2}), and the definition of the phase difference $\delta_{21}\equiv \phi_2-\phi_1$, 
this equation can be rewritten as 
\begin{eqnarray}
E_1^2(\langle \mathcal{K}_3\rangle_\alpha-\langle \mathcal{K}_0\rangle_\alpha)+E_2^2(\langle \mathcal{K}_3\rangle_\alpha+\langle \mathcal{K}_0\rangle_\alpha)\nonumber\\
-E_1 E_2\sqrt{\langle \mathcal{K}_1\rangle_\alpha^2+\langle \mathcal{K}_2\rangle_\alpha^2}\cos\delta_{21}
=2(\langle \mathcal{K}_1\rangle_\alpha^2+\langle \mathcal{K}_2\rangle_\alpha^2)\sin^2\delta_{21}\label{ELIPSE}.
\end{eqnarray}
In this way we have written the polarization ellipse in terms of the classical values of the $su(1,1)$ Lie algebra 
generators.\\
 
We define the classical Stokes-like parameters $K_i$ as follows 
\begin{eqnarray}
&&K_0= \langle \mathcal{K}_0\rangle_\alpha, \hspace{8ex}K_1=\sqrt{\langle \mathcal{K}_1\rangle_\alpha^2+\langle \mathcal{K}_2\rangle_\alpha^2}\sin{\delta_{21}},\label{stokes1}\\
&&K_2=\sqrt{\langle \mathcal{K}_1\rangle_\alpha^2+\langle \mathcal{K}_2\rangle_\alpha^2}\cos{\delta_{21}}, \hspace{5ex}K_3=\langle \mathcal{K}_3\rangle_\alpha.\label{stokes2}
\end{eqnarray}
Thus, the ellipse polarization can be expressed as
\begin{equation}
E_1^2(K_3-K_0)+E_2^2(K_3+K_0)-2E_1 E_2 K_2= 2 K_1^2.
\end{equation}
We note that the polarization  ellipse, equation (\ref{ELIPSE}), depends on the phase difference $\delta_{21}$, whereas the 
classical limit of the $su(1,1)$  generators do not.  This fact has forced us to define the Stokes-like parameters as those of equations (\ref{stokes1}) and (\ref{stokes2}), which depend on the phase difference and are time-independent.        

\section{Stokes-like parameters and the Poincar\'e hyperboloid}

The procedure to map the standard Stokes parameters to the Poincar\'e sphere is explained in full detail in   
Sec. 1.4.2 of Ref. \cite{born}. In the present section, we translate the Born and Wolf ideas to our Stokes-like parameters.
We shall show that they are mapped to a one-sheet of a two-sheet hyperbolic paraboloid. 
       
We perform an hyperbolic rotation to the fields $E_1$  and $E_2$ by an angle $\psi$, in such a way that 
the fields $E_\xi$ and $E_\eta$ are obtained on the principal axis of the ellipse. Thus,
\begin{eqnarray}
E_\xi= E_1 \cosh\psi+ E_2 \sinh\psi,\label{rot1}\\
E_\eta=E_1\sinh\psi+E_2\cosh\psi. \label{rot2}
\end{eqnarray}
Equation (\ref{campo}) can be written as 
\begin{eqnarray}
\vec{E}(z,t)&&=2|\alpha_{01}|\cos(\omega t-kz-\phi_1)\mathbf{i}+2|\alpha_{02}|\cos(\omega t-kz-\phi_2)\mathbf{j}\nonumber \\
                    &&=\tilde{a_1}\cos(\tau-\phi_1)\mathbf{i}+\tilde{a}_2\cos(\tau-\phi_2)\mathbf{j}=E_1\mathbf{i}+E_2\mathbf{j},\label{oscilaciones} 
\end{eqnarray}
where  $\tau=\omega t-kz$, and $\tilde{a_i}=2|\alpha_{0i}|$, $i=1,2$. \\

On the other hand, the oscillations on the principal axis of the ellipse are given by
\begin{eqnarray}
E_\xi=a\cos(\tau+\delta_0),\label{oscilaxi}\\
E_\eta=\pm b \sin(\tau+\delta_0)\label{oscilaeta}.
\end{eqnarray}
being $\delta_0$ a constant phase. The signs $\pm$ are for right and left polarization on the coordinates
system $\xi-\eta$. From the substitution of equations (\ref{oscilaciones}) and (\ref{oscilaxi}) into equation (\ref{rot1}), we deduce the equalities
\begin{eqnarray}
&&a\cos\delta_0=\tilde{a}_1\cosh\psi \cos\phi_1+\tilde{a}_2\sinh\psi\cos\phi_2,\label{seis}\\
&&a\sin\delta_0=-\tilde{a}_1\cosh\psi \sin\phi_1-\tilde{a}_2\sinh\psi \sin\phi_2.\label{siete}
\end{eqnarray}
Similarly, by substituting  equations (\ref{oscilaciones}) and (\ref{oscilaeta}) into equation (\ref{rot2}), we deduce
\begin{eqnarray}
&&\pm b\cos\delta_0=\tilde{a}_1\sinh\psi \sin\phi_1+\tilde{a}_2\cosh\psi\sin\phi_2,\label{ocho}\\
&&\pm b\sin\delta_0=\tilde{a}_1\sinh\psi \cos\phi_1+\tilde{a}_2\cosh\psi\cos\phi_2.\label{nueve}
\end{eqnarray}
Dividing (\ref{ocho}) by (\ref{seis}) and (\ref{nueve}) by  (\ref{siete}), we obtain
\begin{eqnarray}
\pm \frac{b}{a}&=&\frac{\tilde{a}_1\sinh\psi \sin\phi_1+\tilde{a}_2\cosh\psi\sin\phi_2}{\tilde{a}_1\cosh\psi \cos\phi_1+\tilde{a}_2\sinh\psi\cos\phi_2}\nonumber\\
&=&\frac{\tilde{a}_1\sinh\psi \cos\phi_1+\tilde{a}_2\cosh\psi \cos\phi_2}{-\tilde{a}_1\cosh\psi \sin\phi_1-\tilde{a}_2\sinh\psi \sin\phi_2}.
\end{eqnarray}
The second equality and some elementary hyperbolic identities  lead to
 \begin{equation}
\tanh(2\psi)=-\frac{2\tilde{a}_1\tilde{a}_2}{\tilde{a}_1^2+\tilde{a}_2^2} \cos\delta_{21}.\label{funda3}
\end{equation}
The definition $\tanh\alpha\equiv-\frac{\tilde{a}_2}{\tilde{a}_1}$ and the
identity for $\tanh(2\alpha)$ permit to write this equation as   
\begin{equation}
\tanh(2\psi)=\tanh(2\alpha)\cos\delta_{21}
\end{equation}

On squaring and adding (\ref{seis}) and (\ref{siete}), we obtain
\begin{equation}
a^2=\tilde{a}_1^{2}\cosh^2\psi+ \tilde{a}_2^2\sinh^2\psi+2\tilde{a}_1\tilde{a}_2\sinh\psi\cosh\psi \cos\delta_{21}.\label{acuad} 
\end{equation}
Also, on squaring and adding (\ref{ocho}) and (\ref{nueve}):
\begin{equation}
b^2=\tilde{a}_1^{2}\sinh^2\psi+ \tilde{a}_2^2\cosh^2\psi+2\tilde{a}_1\tilde{a}_2\sinh\psi\cosh\psi \cos\delta_{21}.\label{bcuad} 
\end{equation}
The  subtracting of equation (\ref{bcuad}) from equation (\ref{acuad}) allows to obtain
\begin{equation}
a^2-b^2=\tilde{a}_1^2- \tilde{a}_2^2
\end{equation}
If we multiply (\ref{seis}) by (\ref{ocho}), and  (\ref{siete}) by (\ref{nueve}) and adding the respective results, we get
\begin{equation}
\mp ab=\tilde{a}_1\tilde{a}_2\sin\delta_{21}
\end{equation}
From the last two equalities it is implied that 
\begin{equation}
\pm \frac{2ab}{a^2-b^2}=-\frac{2\tilde{a}_1\tilde{a}_2}{\tilde{a}_1^2-\tilde{a}_2^2}\sin\delta_{21}.
\end{equation}
If we introduce the definition $\tanh\chi\equiv \pm \frac{b}{a}$, this equation result to be 
\begin{equation}
\sinh(2\chi)=-\frac{2\tilde{a}_1\tilde{a}_2}{\tilde{a}_1^2-\tilde{a}_2^2}\sin\delta_{21}.\label{funda4}
\end{equation}  
 Now, the definition $\tanh\alpha=-\frac{\tilde{a}_2}{\tilde{a}_1}$ above,  leads to 
\begin{equation}
\sinh(2\chi)=\sinh(2\alpha)\sin\delta_{21}
\end{equation}

Equation (\ref{funda4}) can be written in terms of equations (\ref{funda1}), (\ref{funda2}) and (\ref{oscilaciones}) as
\begin{equation}
\sinh(2\chi)=-\frac{\sqrt{ \langle \mathcal{K}_1\rangle_\alpha^2+\langle \mathcal{K}_2\rangle_\alpha^2}\sin\delta_{21}}{\langle \mathcal{K}_0\rangle_\alpha}=-\frac{K_1}{K_0},
\end{equation}
where for the last equality, we have used equation (\ref{stokes1}). This equation leads to
\begin{equation}
K_1=-K_0\sinh(2\chi),
\end{equation}
or depending on the sign of the $K_0$:
\begin{equation}
K_1=\pm |K_0|\sinh(2\chi).\label{tild1}
\end{equation}
Similarly, from equations (\ref{funda3}),  (\ref{funda1}), (\ref{funda2}),  and (\ref{stokes2}), we 
can show that 
\begin{equation}
K_2=-K_3\tanh(2\psi).\label{tild2}
\end{equation}
On the other hand, from  equations (\ref{funda1}) and (\ref{funda2}) it is immediate to show that 
\begin{equation}
\langle \mathcal{K}_3\rangle_\alpha^2-\langle \mathcal{K}_0\rangle_\alpha^2=\langle \mathcal{K}_1\rangle_\alpha^2+\langle \mathcal{K}_2\rangle_\alpha^2,\hspace{3ex}\hbox{or}\hspace{3ex}
K_3^2-K_0^2=K_1^2+K_2^2,\label{forc}
\end{equation}
where for the last equality we have used equations (\ref{stokes1})  and (\ref{stokes2}). Equations
(\ref{tild1}), (\ref{tild2}) and  (\ref{forc}) lead to
\begin{equation}
K_3=\pm |K_0|\cosh(2\chi)\cosh(2\psi).
\end{equation}
However, because of equation (\ref{stokes2}), $K_3\ge 0$, thus  
\begin{equation}
K_3=+|K_0|\cosh(2\chi)\cosh(2\psi).\label{tild3}
\end{equation}
Substituting this equation into equation (\ref{tild2}) we have
\begin{equation}
K_2=\pm |K_0|\cosh(2\chi)\sinh(2\psi)\label{tild4}.
\end{equation}

Equations (\ref{tild1}),  (\ref{tild3})  and (\ref{tild4}) are the Stokes-like parameters written in terms 
of $\chi$ and  $\psi$. We have plotted the 
vector function 
\begin{eqnarray}
H_p=(\pm |K_0|\sinh(2\chi), \pm |K_0|\sinh(2\chi)\cosh(2\psi),\nonumber\\
+|K_0|\cosh(2\chi)\cosh(2\psi)) 
\end{eqnarray}
with the different setting of signs: a) (+,+,+), b) (+,--,+), c) (--,+,+) and d) (--,--,+).  It must be emphasized that
for each of these setting, the polarization surface generated for the Stokes-like parameters results to be as that 
shown in Fig. 1, been $|K_0|$ the distance from the origin of coordinates to the minimum of the hyperboloid sheet.
Hence, because of the plus sign in equation (\ref{tild3}), the Stokes-like parameters generate a one-sheet 
(Poincar\'e hyperboloid) of a two-sheet hyperboloid, instead an sphere. 
Fig. 1 shows how the Poincar\'e hyperboloid varies as a function of $|K_0|$.  \\

\begin{figure}[h]
\begin{center} 
\includegraphics[scale=0.7]{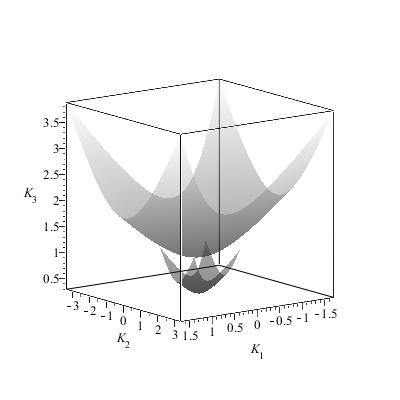}
\caption{It shows how the Poincar\'e hyperboloid changes as a function of $|K_0|$. The greater one is for $|K_0|=1$ and 
the little one is for  $|K_0|=0.3$. }
\end{center}\label{f1}
\end{figure}
 Now, we are interested in investigating the regions on the Poincar\'e hyperboloid in which occurs the
different polarizations: I)linear polarization $LP$, II)circular polarization $CP$ and III) elliptical polarization $EP$. \\

I) For linear polarization $LP$, $\delta_{21}=m\pi$, $m=0,\pm 2, \pm 4, ...$ 
\cite{hecht,born}, equations (\ref{stokes1}) and (\ref{stokes2}) imply that $K_1=0$ and $K_2=$  to a real quantity.
Also, for  $\delta_{21}=m\pi$, $m=\pm 1, \pm 3, \pm 5, ...$  a linear polarization holds. For this case, $K_1=0$ and $K_2=$  to a  
real quantity. Geometrically speaking, the straight line with $K_1=0$ is equal to the $K_2$-axis. Therefore, the projection  of the $K_2$-axis on the one-sheet hyperboloid represents the curve where $LP$ occurs, as it is shown in Fig. 2.  \\

II) The circular polarization $CP$, occurs when $\delta_{21}=\pm\frac{\pi}{2}+2m\pi$,  
( $\pm$  represents right (left) circular polarization, respectively), $m=0,\pm 1, \pm 2,...$ and $|\alpha_{01}|=|\alpha_{02}|$.  In order to have a graphical interpretation to this case, we consider a limit process $|\alpha_{01}|\rightarrow |\alpha_{02}|$  and $|\alpha_{01}|\ne|\alpha_{02}|$. Under this assumption: a) the minimum of the Poincar\'e hyperboloid approaches to the origin of coordinates, as shown in Fig. 1, and b) with $\delta_{21}=\pm\frac{\pi}{2}+2m\pi$, equations (\ref{funda1}), (\ref{funda2}), (\ref{stokes1}) and (\ref{stokes2}) implies that $K_1=\pm |\alpha_{01}||\alpha_{02}|$  and $K_2=0$. In fact, for right circular polarization, $\delta=\frac{\pi}{2}+2m\pi$, $K_1> 0$ and $K_2=0$, and for left circular polarization, $\delta=-\frac{\pi}{2}+2m\pi$, $K_1< 0$ and $K_2=0$. Thus, circular polarization is characterized by $K_2=0$. In this case,  the straight line $K_2=0$  coincides with the $K_1$-axis. Hence, as it is shown in Fig. 2, the projection of the $K_1$-axis 
on the one-sheet hyperboloid represents the points where $CP$ occurs. \\

III) It is well known that right elliptical polarization $REP$ holds for $ 0<\delta_{21}< \pi$, and left 
elliptical polarization $LEP$  holds for $\pi < \delta_{21}<2\pi$ \cite{hecht,born}. Since for $REP$, $\sin\delta_{21}> 0$, from
equations (\ref{stokes1}) and (\ref{stokes2}) we deduce that $K_1> 0$  whereas $K_2$ can takes positive and negative values. Similarly, for $LEP$, $K_1< 0$ and $K_2$ can takes both positive and negative values.  In Fig. 2, if we look at the $K_1-K_2$ plane as the one-sheet hyperboloid domain, then, the $REP$ region is that on the hyperboloid which is on the half-plane for $K_1> 0$.  Similarly, the hyperboloid surface on the half-plane for $K_1<0$,  corresponds to the region where $LEP$ occurs.          
\begin{figure}[h]
\begin{center} 
\includegraphics[scale=0.7]{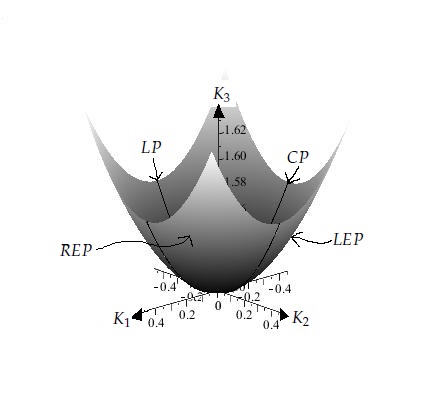}
\caption{One-sheet Poincar\'e hyperboloid. It is plotted the surface generated by the vector function 
$H_p=(+|K_0|\sinh(2\chi),+|K_0|\sinh(2\chi)\cosh(2\psi),+|K_0|\cosh(2\chi)\cosh(2\psi)) $.  We have set 
$K_0=\frac{3}{2}$, $ -0.4\le 2\chi\le 0.4$ and $ -0.4\le 2\psi\le 0.4$. The curve $LP$ represents the points where linear polarization
occurs, and in each point on the curve $CP$ there exists a circular polarization. Each point on the region $REP$ represents a right elliptical polarization, and the points on the
region $LEP$ are those on which left elliptical polarization holds.  }
\end{center}\label{f2}
\end{figure}

\section{Concluding Remarks}

We have given an alternative definition of the Stokes parameters (Stokes-like parameters)
and showed  that the polarization ellipse can be written in terms of these parameters.
Based on the procedure to map the standard Stokes parameters to the Poincar\'e sphere \cite{born},
we achieved to map our Stokes-like parameters 
to a one-sheet (Poincar\'e hyperboloid) of a two-sheet hyperboloid. We note that in the equations  
(\ref{stokes1}) and (\ref{stokes2}), the  
definitions of $K_1$  and $K_2$ can be interchanged without any essential change in the results.

The standard Stokes parameters have been deduced from the quantum regime from the $su(2)$ Lie algebra, whereas the so-called generalized Stokes parameters have been deduced  from the  Kemmer algebra and the $su(3)$ Lie algebra. In each of these algebras, all the corresponding Lie algebra generators commute with the Hamiltonian of the two- or three-dimensional harmonic oscillator (the algebra is called  a symmetry algebra). The existence of a symmetry algebra is the responsible that at the classical limit the phase difference emerge naturally in the standard or in the generalized Stokes parameters \cite{mota,canadian}.  
This result (although it was not in the modern language) between observables (symmetry algebra) and the standard Stokes parameters 
was discovered many years ago by Wolf \cite{wolf}. However, in the present work we have used the $su(1,1)$ Lie algebra, being the Hamiltonian of the two-dimensional 
harmonic oscillator one of its three generators. This feature has as consequence that at the classical
limit the $su(1,1)$ algebra generators (Equations  (\ref{funda1}) and (\ref{funda2})) do not depend on the phase difference.  Thus, the use of the non-compact Lie algebra is the reason why in our definition of the Stokes-like parameters we have attached them a phase difference.
However, in spite of this, the results reported in the present paper formally arise in a natural way and are the fundamental ones to be described by Stokes parameters.

Finally, we emphasize that the results of the present paper are under the assumption that the
direction of arrival  of the electromagnetic wave is known a priori. When the 
direction of arrival of the electromagnetic wave is unknown a priori, a generalization 
of our results could related with the generators of the 
$so(3,1)$ Lie algebra. Applications  of our definition to describe the polarization process in particular optical systems is a work in progress.

\section*{Acknowledgments}
This work was partially supported by SNI-M\'exico, COFAA-IPN, EDI-IPN, EDD-IPN and
CGPI-IPN Project Numbers 20161727 and 20160108.


\begin{thebibliography}{99}

\bibitem{stokes}G. G.Stokes, On the composition and resolution of streams of
polarized light from different sources, Trans. Cambridge Philos., \textbf{9},
399 (1852).

\bibitem{sur}W. A.Shurcliff and S. S.Ballard, Polarized Light, 
Van Nostrand Company, Princeton, N.J., 1964.

\bibitem{mc1} W. H. McMaster, Polarization and the Stokes Parameters,
Am. J. Phys. {\bf 22}, 351(1954).

\bibitem{mc2}W. H.McMaster, Matrix Representation of Polarization, 
Rev. Mod. Phys. {\bf 33}, 8(1961).

\bibitem{colet}E.Collett,The Description of Polarization in Classical Physics, 
Am. J. Phys. {\bf 36}, 713(1968).

\bibitem{wolf} E.Wolf, Optics in Terms of Observable Quantities,
Nuovo Cimento {\bf 12}, 884(1954).

\bibitem{wolf59} E. Wolf,Coherence Properties of partially Polarized Electromagnetic Radiation,
Nuovo Cimento, {\bf 13}, 1165(1959).

\bibitem{blaskal-kim} S.Blaskal  and Y. S.Kim, Symmetries of the Poincar\'e Sphere and Decoherence Matrices, arXiv:quant-ph/0501050. 

\bibitem{roman}P. Roman, Generalized Stokes Parameters for Waves with Arbitrary Form,
Nuovo Cimento {\bf 13}, 974(1959).

\bibitem{rama} G. Ramachandran, M. V. N. Murthy and K. S. Mallesh, SU(3) Representation for the Polarisation of Light, 
Pramana {\bf 15}, 357(1980).

\bibitem{carozzi}T. Carozzi, R. Karlsson  and J. Bergman, Parameters Characterizing Electromagnetic Wave Polarization,
Phys. Rev. E {\bf 65}, 2024(2000).

\bibitem{mota} R. D. Mota, M. A. Xicot\'encatl and V. D. Granados, Jordan-Schwinger map, 3D Harmonic 
Oscillator Constants of Motion, and Classical and Quantum Parameters Characterizing Electromagnetic Wave
Polarization, J. Phys A: Math.and Gen. {\bf 37}, 2835(2004).

\bibitem{jauch}J. M. Jauch and F. Rohrlich, The Theory of
Photons and Electrons, Springer-Verlag, Berlin, 1976.

\bibitem{canadian}R. D. Mota, M. A. Xicot\'encaltl and V. D. Granados, 2D Isotropic Harmonic Oscillator  Approach to the Classical and Quantum Stokes Parameters, Can. J. Phys. {\bf 82}, 767(2004).

\bibitem{sergienko}A. F. Abouraddy, A. V. Sergienko, B. E. A. Saleh and M. C. Teich, Quantum Entanglement and the Two-Photon Parameters, Opt. Commun.  {\bf 201}, 93(2002).

\bibitem{jaeger}G. Jaeger, M. Teodorescu-Frumosu, A. B. Sergienko, B. E. A. Saleh
and M. C. Teich, Multiphoton Stokes-Parameter Invariant for Entangled States, 
Phys. Rev. A {\bf 67}, 032307--(2003).

\bibitem{bieden} L. C. Biedenharn and J. D.Louck, Angular Momentum in
Quantum Physics, Addison-Wesley Publishing Company, Massachusetts, 1981.

\bibitem{cohen}C.Cohen-Tannoudji, B.Diu and F. Lal\"oe, Quantum Mechanics, 
John Wiley $\&$ Sons, Paris, 1977.

\bibitem{guenther}R.Guenther, Modern  Optics, John Wiley $\&$ Sons, New York, 1990.

\bibitem{hecht}E.Hecht, Optics, Addison Wesley, San Francisco, 2002.

\bibitem{born}  M. Born and E. Wolf, Principles of Optics,
Pergamon Press, Oxford, 1975.

\end{thebibliography}
\end{document}